\documentclass[fleqn,11pt,twoside]{article}

\usepackage{amsthm,amsthm,amssymb, color, xcolor,epsfig, graphics, subfigure}

\usepackage{amsmath, graphicx, latexsym, lscape}

\makeatletter
\newcommand{\copyrightnote}[2]{{\renewcommand{\thefootnote}{}
 \footnotetext{\small\it
\begin{flushleft}
 \copyright \ #1   #2  
\end{flushleft}}}}

\newcommand{\Name}[1]{\begin{flushleft}
                       \LARGE \bf #1
                       \end{flushleft}\vspace{-3mm}}

\newcommand{\Author}[1]{\begin{flushleft}
                       \it #1 \end{flushleft}}

\newcommand{\Address}[1]{\begin{flushleft}
                       \it #1 \end{flushleft}}

\newcommand{\Date}[1]{\begin{flushleft}
                      \small  \it #1 \end{flushleft}}

\newcommand{\evenhead}{Author \ name}
\newcommand{\oddhead}{Article \ name}

\renewcommand{\@evenhead}{
\hspace*{-3pt}\raisebox{-15pt}[\headheight][0pt]{\vbox{\hbox to \textwidth
{\thepage \hfil \evenhead}\vskip4pt \hrule}}}
\renewcommand{\@oddhead}{
\hspace*{-3pt}\raisebox{-15pt}[\headheight][0pt]{\vbox{\hbox to \textwidth
{\oddhead \hfil \thepage}\vskip4pt\hrule}}}
\renewcommand{\@evenfoot}{}
\renewcommand{\@oddfoot}{}

\long\def\@makecaption#1#2{%
  \vskip\abovecaptionskip
  \sbox\@tempboxa{\small \textbf{#1.}\ \ #2}%
  \ifdim \wd\@tempboxa >\hsize
    {\small \textbf{#1.}\ \ #2}\par
  \else
    \global \@minipagefalse
    \hb@xt@\hsize{\hfil\box\@tempboxa\hfil}%
  \fi
  \vskip\belowcaptionskip}

\newcommand{\JNMPnumberwithin}[3][\arabic]{%
  \@ifundefined{c@#2}{\@nocounterr{#2}}{%
    \@ifundefined{c@#3}{\@nocnterr{#3}}{%
      \@addtoreset{#2}{#3}%
      \@xp\xdef\csname the#2\endcsname{%
        \@xp\@nx\csname the#3\endcsname .\@nx#1{#2}}}}%
}

\newcommand{\resetfootnoterule} {
  \renewcommand\footnoterule{%
  \kern-3\p@
  \hrule\@width.4\columnwidth
  \kern2.6\p@}
}

\renewcommand{\footnoterule}{}

\makeatother

\theoremstyle{definition}

\newcommand{\beq}{\begin{equation}}  
\newcommand{\eeq}{\end{equation}}  
\newcommand{\bea}{\begin{eqnarray}} 
\newcommand{\eea}{\end{eqnarray}}   
\newcommand{\bear}{\begin{array}}  
\newcommand{\eear}{\end{array}}

\newtheorem{thm}{Theorem}[section] 

\newtheorem{propn}[thm]{Proposition}

\newenvironment{prf}{\trivlist \item [\hskip 
\labelsep {\bf Proof:}]\ignorespaces}{\qed \endtrivlist}

\newcommand{\Z}{{\mathbb Z}}
\newcommand{\C}{{\mathbb C}}

\newcommand{\rd}{\mathrm{d}}

\newcommand\al{{\alpha}}

\newcommand\de{{\delta}}

\newcommand\lax{{\bf L}}
\newcommand\mma{{\bf M}}

\newcommand{\cQ}{{\cal Q}}
\newcommand{\cP}{{\cal P}}

\newcommand{\cR}{{\cal R}}

\begin{document}

\renewcommand{\evenhead}{ {\LARGE\textcolor{blue!10!black!40!green}{{\sf \ \ \ ]ocnmp[}}}\strut\hfill  A N W Hone, 
J A G Roberts, P Vanhaecke and F Zullo}
\renewcommand{\oddhead}{ {\LARGE\textcolor{blue!10!black!40!green}{{\sf ]ocnmp[}}}\ \ \ \ \  Integrable maps in 4D and modified Volterra lattices}

\thispagestyle{empty}
\newcommand{\FistPageHead}[3]{
\begin{flushleft}
\raisebox{8mm}[0pt][0pt]
{\footnotesize \sf
\parbox{150mm}{{Open Communications in Nonlinear Mathematical Physics}\ \ \ \ {\LARGE\textcolor{blue!10!black!40!green}{]ocnmp[}}
\quad Special Issue 1, 2024\ \  pp
#2\hfill {\sc #3}}}\vspace{-13mm}
\end{flushleft}}

\FistPageHead{1}{\pageref{firstpage}--\pageref{lastpage}}{ \ \ }

\strut\hfill

\strut\hfill

\copyrightnote{The author(s). Distributed under a Creative Commons Attribution 4.0 International License}

\begin{center}
{  {\bf This article is part of an OCNMP Special Issue\\ 
\smallskip
in Memory of Professor Decio Levi}}
\end{center}

\smallskip

\Name{Integrable maps in 4D and modified Volterra lattices}

\Author{A.N.W. Hone$^{\,1}$,  J.A.G. Roberts$^{\,2}$, P. Vanhaecke$^{\,3}$ and F. Zullo$^{\,4}$}

\Address{$^{1}$ School of Mathematics, Statistics \& Actuarial Science,
    University of Kent\\ Canterbury CT2 7NF, UK\\[2mm]
$^{2}$ School of Mathematics \&  Statistics, University of New South Wales, 
  Sydney\\ NSW 2052, Australia\\[2mm]
$^{3}$ Laboratoire de Math\'ematiques 
\&  Applications, UMR 7348~CNRS - Universit\'e de Poitiers, 
  86360 Chasseneuil-du-Poitou, France\\[2mm]
$^{4}$ DICATAM, Universit\`a degli Studi di Brescia,  via Branze 38 - 
25123 Brescia\\ 
\&  INFN, sezione di Milano-Bicocca, 
 Piazza della Scienza 3 - 20126, Milano, Italia}

\Date{Received October 31, 2023; Accepted February 12, 2024}

\setcounter{equation}{0}

\begin{abstract}

\noindent 
In recent work, we presented the construction of a family of difference equations associated with the Stieltjes continued fraction expansion of a certain function on a hyperelliptic curve of genus $g$. As well as proving that each such discrete system is  an integrable map in the Liouville sense, we also showed it to be an
algebraic completely integrable system. In the discrete setting, the latter means that the generic level set of the invariants 
is an affine part of an abelian variety, in this case the Jacobian of the hyperelliptic curve, and  each iteration of the map corresponds to a translation by a fixed vector on the Jacobian. 
In addition, we demonstrated that, by combining 
the discrete integrable dynamics with the flow of one of the commuting Hamiltonian vector fields, these maps provide genus $g$ algebro-geometric solutions of the infinite 
Volterra lattice, which justified naming them \textit{Volterra maps}, denoted ${\cal V}_g$. 
 
The original motivation behind our work was the fact that, 
in the particular case $g=2$, we could recover an example of an integrable symplectic map in four dimensions  
found by Gubbiotti, Joshi, Tran and Viallet, who classified birational maps 
in 4D admitting two invariants
(first integrals) with a particular degree structure,  by considering recurrences of 
fourth order with a certain symmetry. 
Hence, in this particular case, the map ${\cal V}_2$ yields genus two solutions of the Volterra lattice. 
The purpose of  this note is to  point out how two of the other 4D integrable maps obtained in the classification 
of Gubbiotti et al.\ correspond to genus two solutions of two different forms of the modified Volterra lattice, 
being related via  
a Miura-type transformation to the $g=2$ Volterra map ${\cal V}_2$. 

We  dedicate this work to a dear friend and colleague, Decio Levi. 
\end{abstract}

\label{firstpage}


\section{Introduction}

This short article consists of some recollections of our colleague Decio Levi (in section 2 below), followed 
by a brief update on our recent results about integrable maps in four (and higher) dimensions, 
which provide algebro-geometric solutions of differential-difference equations of Volterra type  
\cite{voltmaps}. Decio was one of the pioneers in the theory of integrability for  differential-difference equations, 
especially in the construction of  integrable lattices from B\"acklund transformations for continuous systems \cite{levib, levibt}, 
and the programme of applying the symmetry approach to the classification of such lattices, which he initiated with Yamilov \cite{leviy}. 
Thus we like to think that Decio would have appreciated the results being presented here. 

After presenting a few memories of Decio, in section 3 we begin by giving a brief overview of the 4D integrable maps which were classified 
by  Gubbiotti et al.\  \cite{gjtv2}. We then proceed to review our construction of integrable maps obtained from the Stieltjes fraction expansion  
of certain functions on hyperelliptic curves \cite{voltmaps}, and explain how it reproduces one of the examples from \cite{gjtv2}, denoted (P.iv), 
in the particular case of genus two curves. Sections 4 and 5 are devoted to the maps (P.v) and (P.vi), respectively: we show how each of 
these maps is related to a different form of the modified Volterra lattice, and present explicit formulae which relate their solutions to 
the solutions of (P.iv) via a transformation of Miura type. We end with some very short conclusions in section 6.

\section{Memories of Decio Levi} 
\textit{Andrew Hone writes: } 
I first met Decio in Warsaw in September 1995, when I was a PhD student participating in the 1st Non-Orthodox School on
Nonlinearity and Geometry \cite{nonorth}. Decio was one of the lecturers, along with Orlando Ragnisco, and it was 
thanks to extended conversations with Orlando that I resolved to apply for postdoctoral  funding to work with him when 
I finished my PhD. After receiving a grant from the Leverhulme Trust two years later, I finally  got to be a researcher at Roma Tre, where Orlando and Decio 
were both professors in the Dipartimento di Fisica. 

For approximately the first six months of my time in Rome, there was no available office space for postdocs, which meant that I had to share an office with Orlando. 
Far from being a negative aspect of my experience, this situation had many positive benefits for me, and not just scientific ones. By working in close proximity with 
Orlando, it meant that I was privy to the regular visits from the neighbour in the office next door, namely Decio, his long-time friend and collaborator. Apart from 
the pleasure of getting to know Decio, and learning many wonderful ideas about integrable systems from him, there was the fact that, 
by default, he would chat to Orlando in Italian,  
which helped me to rapidly improve my grasp of the language in those first few months. 
The strong bond of friendship between Orlando and Decio created a very happy atmosphere, 
and I have extremely fond memories of those times. 

In subsequent years, I would see Decio fairly often at various international conferences, or during return visits to Rome. 
He had an amiable manner and a warm, cheerful smile. It was always enjoyable to talk to him, whether about technical problems, 
sharing family news, or just musing about life in general. Talking with Decio would leave me feeling reassured, that all was right 
with the world, and I liked his gentle way of concluding a long conversation with ``Vabb\`e in somma''. 

It is an honour to be able to remember Decio here, both for his contributions as a scientist, and as a wonderful human being. \\~\\
\textit{Federico Zullo writes: } 
The first time I met Decio was in 2003: I was a student at the Dipartimento di Fisica of Roma Tre University and needed an advisor for my last examination for my laurea triennale (bachelor's degree). I asked Orlando Ragnisco who, at that time, was very busy. He accompanied me to the office next door, 
where Decio was, and I asked him for a theme for my short dissertation. He very heartily introduced me to the subject of solitons, that I never heard about before, giving me books and kind advice. Later, during my laurea magistrale (master's degree), and during my PhD studies, I followed different classes taught by Decio, some with very few students. The familiar atmosphere and natural mildness of Decio's classes fostered my learning, and I'm greatly indebted to him for having taught me many topics used in mathematical physics, like group theory, symmetries of differential equations, physics of nonlinear systems, qualitative and quantitative analysis of solutions of differential equations and others. 
For my own teaching, I still use some of the material that I collected from his courses. 

For a period just before 2014, I was hosted by Decio in his office as a researcher. 
I remember the talks on disparate subjects, like religion, literature, politics, society and, obviously, our research. The talks would then continue during the lunch break, usually in Via Marconi, with Orlando and the other members of the very stimulating group of young researchers that was gathered at Roma Tre in that period, including Fabio Musso, Matteo Petrera, Christian Scimiterna, Danilo Riglioni, Riccardo Droghei, and later Giorgio Gubbiotti and Danilo Latini, all led by Decio and Orlando. I'll always keep these beautiful memories with me.

\section{The map (P.iv) and the geometry of its solutions} 
\setcounter{equation}{0} 

Discrete integrable systems can be constructed by applying an appropriate discretization procedure to continuous ones, and historically this is how 
many examples of discrete integrability were first discovered \cite{levib, suris}. However, from both 
a theoretical point of view and a practical one, it is important to have a notion of integrability for discrete systems that  does not require 
making reference to some underlying continuous system, whether this be for lattice equations \cite{leviy}, or for integrable maps \cite{bruschi, maeda, veselov}. 
While integrable maps in two and three dimensions lead to families of invariant curves (as the level sets of first integrals), the case of four dimensions can lead 
to new features, namely invariant tori of dimension two.  

In \cite{gjtv2}, Gubbiotti et al.\ presented a classification of four-dimensional birational maps of recurrence type, that is 
\beq\label{themap} 
\varphi: \qquad (w_0,w_1,w_2,w_3)\mapsto \Big(w_1,w_2,w_3 ,F(w_0,w_1,w_2,w_3)\Big), 
\eeq 
for a suitable rational function $F$ of  the  affine coordinates  $(w_0,w_1,w_2,w_3)\in \C^4$,  where the map $\varphi$ is required to be  
invariant under the involution $\iota:\, (w_0,w_1,w_2,w_3)\mapsto (w_3,w_2,w_1,w_0)$,  and to possess two independent  polynomial invariants, $H_1$, $H_2$ say,  with specific 
degree patterns $(\deg_{w_0}H_j, \deg_{w_1}H_j, \deg_{w_2}H_j, \deg_{w_3}H_j)=(1,3,3,1)$ and $(2,4,4,2)$ 
for $j=1,2$, respectively. The result of this classification  was six maps with parameters, labelled (P.i-vi), together with six
associated maps, denoted (Q.i-vi) respectively. 
Each of the ``Q'' maps  arises from a corresponding ``P'' map,  as a discrete 
integrating factor for linear combinations of the first integrals, so they are dual to one another  in the sense of \cite{quispel}. 

As described previously, first in \cite{jv} and then \cite{gjtv1}, the original motivation for classifying such maps was   
to understand autonomous versions of the fourth-order members of hierarchies of discrete Painlev\'e I/II equations from \cite{cj}; but, aside from the latter connection,  
the ``P'' in this nomenclature has nothing to do with the usual labelling of continuous Painlev\'e equations.  
From our point of view, the most interesting cases are the maps labelled (P.iv), (P.v) and (P.vi), since 
(from Table 1 in \cite{gjtv2}) these are the only ones arising from a discrete variational principle (Lagrangian), leading 
to a non-degenerate Poisson bracket in four dimensions, such that the two first integrals $H_1$, $H_2$ are in involution; this 
means that in the real case the Liouville tori are two-dimensional.
Subsequently, Gubbiotti obtained these 4D integrable maps via an alternative method, by classifying fourth-order difference equations with a discrete Lagrangian structure \cite{gub}. 

Here we begin with the case of (P.iv), which is the birational map given in affine coordinates by the recurrence 
\small 
\beq\label{pivmap} 
\begin{array}{l} 
w_{n+4} w_{n+3} w_{n+2} 
+
w_{n+2} w_{n+1} w_{n} 
+
2w_{n+2}^2( w_{n+3} +w_{n+1})  +w_{n+2}^3 \\ 
+
w_{n+2} (w_{n+3}^2+ w_{n+3} w_{n+1} +w_{n+1}^2) 
+
\nu w_{n+2} (w_{n+3} +w_{n+2}+ w_{n+1})+ 
b w_{n+2} +a =0.
\end{array} 
\eeq 
\normalsize
This map has three essential parameters $a,b,\nu$  (in the formulae from \cite{gjtv2} we have set the parameter $d=1$, which can be achieved by a simple rescaling), and it is of the form  (\ref{themap}), with  $F$ given by 
\small 
$$
-\frac{w_0w_1w_2+w_1w_2w_3+w_1^2w_2+w_2w_3^2+2w_1w_2^2+2w_2^2w_3+w_2^3
+\nu(w_1w_2+w_2w_3+w_2^2)+bw_2+a}
{w_2w_3}; 
$$ 
\normalsize 
this $F$ 
is the rational function of $w_0,w_1,w_2,w_3$ obtained by 
solving for $w_4$ in (\ref{pivmap}) with $n=0$. 

The first integral denoted $I_{\mathrm{low}}^{\mathrm{P}.\mathrm{iv}}$ in 
\cite{gjtv2} is given in affine coordinates by 
\beq\label{pivh1} 
H_1=w_1w_2\Big(w_2w_3+w_0w_1-w_0w_3
+(w_1+w_2)^2 +\nu(w_1+w_2) + b 
\Big)
+a(w_1+w_2).
\eeq
The latter has the degree pattern $(1,3,3,1)$. In particular, it is linear in $w_3$, which implies that, on each three-dimensional level 
set $H_1=h_1=\,\,$const, the map (\ref{pivmap}) reduces to a birational map in three dimensions, given 
by the recurrence 
$$
\begin{array}{rcl} 
w_{n+3}w_{n+2}w_{n+1}(w_{n+2}-w_n) 
+w_{n+2}w_{n+1}^2w_n 
 +w_{n+2}w_{n+1}(w_{n+1}+w_{n+2})^2 &&  \\ 
+\nu \,w_{n+2}w_{n+1}(w_{n+1}+w_{n+2})+b\,w_{n+2}w_{n+1}
+a\,(w_{n+1}+w_{n+2})&=&h_1. 
\end{array} 
$$  
A second independent invariant for (\ref{pivmap}), with degree pattern $(2,4,4,2)$, is given by 
\small 
\beq\label{pivh2} 
\begin{array}{rcl}  
H_2 &=& w_1w_2\left(\begin{array}{c}  
w_0^2w_1 + w_3^2w_2 + w_0w_3(w_1+w_2) 
+w_0(w_2^2+2w_1^2)  +w_3(w_1^2+2w_2^2) \\ +\,3(w_0+w_3)w_1w_2 +(w_1+w_2)^3 \\ 
+\nu\,\,\Big(w_0w_3+(w_0+w_3)(w_1+w_2) +(w_1+w_2)^2\Big) 
+b \,(w_0+w_1+w_2+w_3)
\end{array}  
\right) \\ 
&& +a\,\Big(w_0w_1+w_3w_2+(w_1+w_2)^2\Big) .
\end{array} 
\eeq 
\normalsize
This differs slightly from the second invariant presented in \cite{gjtv2}, 
which is 
$I_{\mathrm{high}}^{\mathrm{P}.\mathrm{iv}}=H_2-\nu H_1$. 

The nondegenerate Poisson bracket between the coordinates, which was obtained  in \cite{gjtv2} by making use of 
a discrete  Lagrangian for (\ref{pivmap}),  is given by 
\small 
\beq\label{pbiv} 
\{ \, w_n,w_{n+1}\,\} =0, \, 
\{ \, w_n,w_{n+2}\,\} = \frac{1}{w_{n+1}}, \, 
\{ \, w_n,w_{n+3}\,\} = - \frac{w_n +2w_{n+1}+
2w_{n+2}+w_{n+3}+\nu}{w_{n+1}w_{n+2}}, 
\eeq
\normalsize
for all $n$. So  (\ref{themap}) is a Poisson map, 
in the sense 
that $\{\, \varphi^*G,\varphi^*H\,\} =\varphi^*\{\,G,H\,\}$ for all functions $G,H$ 
on $\C^4$. 
The two independent invariants given in \cite{gjtv2} are in involution with respect to this bracket, which 
is equivalent to the involutivity of functions (\ref{pivh1}) and (\ref{pivh2}),  that is to say 
$$\{\,H_1,H_2\,\}=0.$$ 
Hence the four-dimensional map defined by 
(\ref{pivmap}) is integrable in the Liouville sense. 

Computing the Hamiltonian vector field for the first flow, generated by $H_1$, we find 
that this takes the form 
\beq\label{volterra} 
\frac{\rd w_n}{\rd t} = w_n (w_{n+1}-w_{n-1})
\eeq 
for $n=1,2$. However, since (\ref{pivmap}) is a Poisson map that commutes with this flow, it follows that the relation (\ref{volterra}) extends to all 
$n\in\Z$. Thus the combined solutions of the map and the flow, which are compatible with one another, generate a sequence of functions 
$\big(w_n(t)\big)_{n\in\Z}$ satisfying (\ref{volterra}), which is the Volterra lattice equation, first considered by Kac and van Moerbeke \cite{kacvm}. Hence, in a certain sense that can be made precise, these 
will turn out to be genus 2 solutions of this lattice hierarchy.  

The  complex geometry of the solutions of the map defined by (\ref{pivmap}) 
is related to a family of hyperelliptic curves of genus 2, given by the Weierstrass quintic
\beq\label{wg2} 
\Gamma: \quad 
y^2=(1+\nu x +bx^2)^2+4a(1+\nu x +bx^2)x^3+4h_1x^4+4(h_2+\nu h_1)x^5. 
\eeq 
On any genus 2 curve $\Gamma$ of the above form, we take the meromorphic function $F$ given by 
\beq\label{Fdef}
F=\frac{y+{\cal P}(x)}{{\cal Q}(x)}=\frac{{\cal R}(x)}{y- {\cal P}(x)}, 
\eeq 
where ${\cal P},{\cal Q},{\cal R}$ are polynomials in the spectral parameter $x$, given by 
\beq\label{PQR} 
{\cal P}(x)=1+p_1x+p_2x^2, \quad 
{\cal Q}(x)=2+q_1x+q_2x^2, \quad 
{\cal R}(x) =r_1x+r_2x^2+r_3x^3,   
\eeq 
which are required to satisfy 
\beq\label{spec}
{\cal P}(x)^2 +{\cal Q}(x){\cal R}(x)=f(x), 
\eeq 
with $f(x)=(1+\nu x +bx^2)^2+4a(1+\nu x +bx^2)x^3+4h_1x^4+4(h_2+\nu h_1)x^5$ 
being the quintic on the right-hand side of (\ref{wg2}). 
Then the key to the construction in \cite{voltmaps} is to expand the function $F$ as a continued fraction of Stieltjes type (S-fraction), that is 
\beq \label{sfrac} 
F=  1-\cfrac{w_1x}{ 1 -\cfrac {w_2x}{ 1-\cfrac{w_3x}{1- \cdots} } } \;,
\eeq 
and by iterating from one line of the fraction to the next we find that we obtain a recurrence for the coefficients $w_j$. More precisely, the non-trivial coefficients of the polynomials (\ref{PQR}) 
are given in terms of $w_j$ and the parameters by 
$$ 
\begin{array}{l} 
p_1=2w_0+\nu, \quad p_2=2w_0(w_1+w_0+w_{-1})+b, \quad 
\tfrac{1}{2}q_1 =w_0+w_1+\nu , \\ 
\tfrac{1}{2}q_2 =
w_0w_{-1}+w_1w_2+(w_1+w_0)^2+\nu(w_0+w_1)+b , \quad 
r_1=-2w_0; 
\end{array}  
$$
there are similar (but slightly more unwieldy) expressions for $r_2$ and $r_3$, which are omitted here, but are easily obtained from the relation (\ref{spec}).   With these identifications, the iteration of the S-fraction (\ref{sfrac}) for $F$ becomes 
precisely the map (P.iv) in terms of the affine coordinates $w_j$, as given by (\ref{pivmap}).

In \cite{voltmaps} it was also shown that each iteration of the continued fraction is equivalent to the discrete Lax equation 
$$ 
\lax(x)\mma(x) = \mma(x)\widetilde\lax(x)\;, 
$$ 
where 
$$
 \lax(x) := \left(\begin{array}{lr} 
    \cP(x) & \cR(x)\\ 
    \cQ(x) & -\cP(x) 
  \end{array}\right)\;,
\qquad 
\mma(x):=
\left(\begin{array}{cc}
  1 & -w_1 x \\ 
  1 & 0 
\end{array}\right). 
$$ 
Furthermore, we found that each generic common level set of the two invariants $H_1,H_2$ 
is isomorphic to an affine part of the Jacobian of the associated spectral curve $\Gamma$ (or rather, of its completion), and each iteration of the map corresponds to a translation on the Jacobian by the divisor class $[(0,-1)-\infty]$. Thus, in addition to being integrable in the Liouville sense, the map  (\ref{pivmap}) is an algebraic completely integrable system, being a discrete analogue of an a.c.i. system (see \cite{amv, vanhaecke}). 

The map (\ref{pivmap}) can also be rewritten in terms of tau functions $\tau_n$, related to $w_n$ 
via
$$ 
w_n = \frac{\tau_{n}\tau_{n+3}}{\tau_{n+1}\tau_{n+2}}. 
$$  
These tau functions satisfy a Somos-9 recurrence, that is 
\beq\label{s9g2}
\al_1\, \tau_{n+9}\tau_{n} 
+ 
 \al_2\, \tau_{n+8}\tau_{n+1} 
+\al_3\, \tau_{n+7}\tau_{n+2} 
+\al_4\, \tau_{n+6}\tau_{n+3} 
+\al_5\, \tau_{n+5}\tau_{n+4}=0, 
\eeq 
with coefficients $\al_j$ that depend on $a,b,\nu$ and the values of $H_1,H_2$ along each orbit of (\ref{pivmap}); for details see Proposition 2.1 in \cite{voltmaps}. 
Note that the general Somos-9 relation is not integrable, so the initial conditions for (\ref{s9g2})  cannot be freely chosen in a space of dimension 9. However, as explained in 
 \cite{hones6},  such higher order Somos equations 
arise  as linear relations between weight 2 products of sequences of Riemann theta functions. 

Using the S-fraction 
(\ref{sfrac}), we were also able to write explicit Hankel determinant formulae for  these tau functions $\tau_n$, analogous to  results for Somos sequences in genus 1 \cite{chang}, and 
other Hankel determinant formulae for solutions of the Volterra lattice \cite{chen}. Furthermore, we found a Miura map relating the solutions of (P.iv) to one of the maps derived from J-fractions in \cite{contfrac}, using the classical method of 
contraction of continued fractions due to Stieltjes \cite{stieltjes} (see also \cite{shohat}), which in this case turned out to provide solutions 
of the infinite Toda lattice. 

In what follows, we will present analogous properties for the maps (P.v) and (P.vi), and point out how they are closely connected to (P.iv). 

 \section{The map (P.v)} 
\setcounter{equation}{0}

The map (P.v) is given by the recurrence 
\beq\label{pvmap} 
\begin{array}{rcl}
w_{n+4} w_{n+3}^2 w_{n+2}^2 
+
w_{n+2}^2 w_{n+1}^2 w_{n} & 
+ & 
w_{n+2}^3 ( w_{n+1} + w_{n+3})^2
\\ 
& + & \tilde{\nu} w_{n+2}^2 ( w_{n+1} + w_{n+3}) 
+\tilde{c}w_{n+2}+\tilde{a} =0, 
\end{array}  
\eeq 
with three essential parameters $\tilde{a},\tilde{c},\tilde{\nu}$ (compared with \cite{gjtv2} we have put tildes 
here to distinguish them from the parameters in (\ref{pivmap}), and rescaled so that  
the parameter $d\to 1$).  

The lowest degree first integral of the map defined by (\ref{pvmap}), with degree 
pattern $(1,3,3,1)$, is given by 
\beq\label{pvh1} 
H_1 = w_3 w_2^3 w_1^2 +   w_2^2 w_1^3w_0- w_3w_2^2w_1^2w_0+ w_2^3w_1^3 
	+\tilde{\nu} w_2^2 w_1^2+ \tilde{c}w_2 w_1 + \tilde{a}(w_2+ w_1),
\eeq  
and this is the same as 
$I_{\mathrm{low}}^{\mathrm{P}.\mathrm{v}}$ in 
\cite{gjtv2}.
Another first integral, with  degree 
pattern $(2,4,4,2)$, is 
\beq\label{pvh2} 
\begin{array}{rcl}
H_2  & = & w_2^2w_1^2 \Big((w_3w_2+w_1w_0+w_2w_1)^2+\tilde{\nu} (w_3+w_1)(w_2+w_0)\Big) \\ 
&& + \tilde{c}w_2 w_1(w_3w_2+w_1w_0+w_2w_1)  
+ \tilde{a}(w_3w_2^2+ w_1^2w_0+w_2^2w_1+w_2w_1^2).
\end{array} 
\eeq  
The second invariant presented in \cite{gjtv2} is 
$I_{\mathrm{high}}^{\mathrm{P}.\mathrm{v}}=H_2-\tilde{\nu} H_1$.

The nondegenerate Poisson bracket between the coordinates is given by 
$$ 
\{ \, w_n,w_{n+1}\,\} =0, \qquad  
\{ \, w_n,w_{n+2}\,\} = \frac{1}{w_{n+1}^2}, 
$$ 
$$ 
\{ \, w_n,w_{n+3}\,\} = - \frac{2(w_n w_{n+1}+w_{n+1}w_{n+2}+w_{n+2}w_{n+3})+\tilde{\nu}}{w_{n+1}^2w_{n+2}^2}.
$$ 
The independent first integrals (\ref{pvh1}) and (\ref{pvh2}) are in involution with respect to this bracket, 
which shows that the map (\ref{pvmap}) is Liouville integrable. 

Computing the Hamiltonian vector field for the first flow, generated by $H_1$, we find 
that this takes the form 
\beq\label{mvolterra1} 
\frac{\rd w_n}{\rd t} = w_n^2 (w_{n+1}-w_{n-1})
\eeq 
for $n=1,2$. However, since the map (\ref{pvmap}) is Poisson and commutes with the flow $\{\cdot , H_1\}$, the equation (\ref{mvolterra1}) holds for all 
$n\in\Z$. Thus the compatible  solutions of the map and the flow together provide a sequence of functions 
$\big(w_n(t)\big)_{n\in\Z}$ which satisfy (\ref{mvolterra1}), which is a degenerate case of the modified Volterra lattice equation \cite{yu}. 

If we make the tau function substitution
\beq\label{pvtau} 
w_n=\frac{\tau_{n}\tau_{n+2}}{\tau_{n+1}^2}
\eeq 
for (P.v), then we find that 
the sequence $(\tau_n)$ satisfies a Somos-8 relation. More precisely, by direct 
computer algebra calculations we can show the following: 
\begin{propn} Whenever 
$w_n$ is a solution of (\ref{pvmap}),  
the sequence $(\tau_n)$ 
satisfies the following Somos-8 recurrence, with coefficients that are functions of 
the Hamiltonians $H_1, H_2$ as in (\ref{pvh1}) and (\ref{pvh2})  above (constant along 
each orbit): 
\beq\label{s8v}
\al_1\, \tau_{n+8}\tau_{n} 
+ 
 \al_2\, \tau_{n+7}\tau_{n+1} 
+\al_3\, \tau_{n+6}\tau_{n+2} 
+\al_4\, \tau_{n+5}\tau_{3} 
+\al_5\, \tau_{n+4}^2=0, 
\eeq 
where the coefficients are given  by 
$$\begin{array}{c} 
\al_1=H_1, \qquad
\al_2=\tilde{a}H_2, \qquad 
\al_3 = \tilde{a}^2H_2-H_1^3, \\  
\al_4 = \tilde{a}\Big(H_2^2+\tilde{\nu} H_1H_2+\tilde{c}H_1^2+\tilde{a}^2H_1\Big), \qquad
\al_5 = -H_1\Big(H_2^2+\tilde{\nu} H_1H_2+\tilde{c}H_1^2+\tilde{a}^2H_1\Big)
.
\end{array} 
$$ 
\end{propn} 

The reader should note that, just as is the case with (\ref{s9g2}), general higher order relations such as (\ref{s8v}) (generic Somos-$k$ for $k\geq 8$) should not be regarded as discrete integrable systems in their own right, since the coefficients are not arbitrary, and the 
initial values cannot be freely chosen in a space of dimension 8. Nevertheless, bilinear relations of this type appear naturally as tau function constraints arising in Hermite-Pad\'e 
approximation problems (see \cite{ds}). 

Let us denote a solution of the Volterra lattice (\ref{volterra}) by $\hat{w}_n$. Then the Miura map from the modified Volterra lattice (\ref{mvolterra1}) takes the form 
\beq\label{miuramapv} 
\hat{w}_n=w_{n+1}w_{n}.  
\eeq 
This Miura map remains valid at the level of the maps (\ref{pivmap}) and (\ref{pvmap}), in the following sense. 

\begin{thm} \label{miurav} 
Let $w_n$ be a solution of (\ref{pvmap}) with parameters $\tilde{a},\tilde{c},\tilde{\nu}$,  lying on the level set $H_1=\tilde{h}_1$, $H_2=\tilde{h}_2$,  of the first integrals 
 (\ref{pvh1}) and  (\ref{pvh2}). 
Then $\hat{w}_n$ given by the Miura map (\ref{miuramapv}) is a solution of (\ref{pivmap}) with parameters 
$$ 
\nu = \tilde{\nu}, \qquad b=\tilde{c}, \qquad a=\tilde{h}_1.
$$ 
Furthermore, on this solution $\hat{w}_n$, 
the values $h_1, h_2$ of the first integrals  (\ref{pivh1}) and  (\ref{pivh2}) for the map  (\ref{pivmap}) 
are given by 
$$ 
h_1=\tilde{h}_2, \qquad h_2= -\tilde{a}^2-\tilde{\nu}\tilde{h}_2-\tilde{c}\tilde{h}_1. 
$$
\end{thm} 

\begin{prf} The first part of this result is verified by substituting the Miura formula (\ref{miuramapv}) directly into (\ref{pivmap}), using (\ref{pvmap}) to eliminate $w_{n+5}$ followed by 
$w_{n+4}$, and then using the formula for $H_1$ in (\ref{pvh1}) to eliminate $w_{n+3}$ on the level set $H_1=\tilde{h}_1$. Analogous calculations, rewriting  (\ref{pivh1}) and  (\ref{pivh2}) in terms 
of $w_n$ satisfying  (\ref{pvmap}) and comparing with $\tilde{h}_2$, the value of the first integral (\ref{pvh2}) for the latter map, yield the above expressions for $h_1,h_2$.  
\end{prf}
It is worth commenting on the meaning of the Miura formula (\ref{miuramapv}), restricted
to this finite-dimensional setting. Given initial data $w_0,w_1,w_2,w_3$ for 
the map (\ref{pvmap}), we can fix  a level set $H_1=\tilde{h}_1$ to write 
$$ 
\hat{w}_0=w_0w_1, \,\hat{w}_1=w_1w_2,\, \hat{w}_2=w_2w_3, \, 
\hat{w}_3=w_3\, G(w_1,w_2,w_3,\tilde{h}_1),
$$ 
for some rational function  $G$, obtained by using the formula (\ref{pvh1})  
for $H_1$ 
to eliminate $w_4$. Similarly, we can use $H_1$ to eliminate $w_0$ above in terms of $w_1,w_2,w_3$ and $\tilde{h}_1$, and after taking resultants we can do further elimination 
to solve for  each of  $w_0,w_1,w_2,w_3$  as algebraic functions of 
$\hat{w}_0,\hat{w}_1,\hat{w}_2,\hat{w}_3$ and $\tilde{h}_1$. So this leads to an explicit inverse 
of  (\ref{miuramapv}), at least in the form of an algebraic correspondence.

\section{The map (P.vi)} 
\setcounter{equation}{0}

The map (P.vi) is given by 
\small 
\beq\label{pvimap} 
\begin{array}{l} 
w_{n+4} (w_{n+3}^2-\de^2)(w_{n+2}^2-\de^2) 
+
w_{n} (w_{n+1}^2-\de^2)(w_{n+2}^2-\de^2) 
\\ 
+ 
w_{n+2}\Big( (w_{n+2}^2-\de^2)(w_{n+3}+w_{n+1})^2+\bar{c}-\de^4\Big)
+\bar{\nu}(w_{n+2}^2-\de^2)(w_{n+3}+w_{n+1}) + \bar{a}  =  0.
\end{array} 
\eeq 
\normalsize
This depends on only three essential parameters $\bar{a},\bar{c},\bar{\nu}$; compared with \cite{gjtv2} 
we have replaced $a\to\bar{a}$, $c\to\bar{c}$, $d\to -\bar{\nu}$ and $\delta\to\delta^2$. Note the map P(v) in the previous section 
arises from P(vi) in the limit $\delta\to 0$, while 
for $\delta\neq 0$ the map can always be rescaled so that $\delta\to 1$, but it will be convenient to retain this parameter which 
has the same weight as $w_n$ in (\ref{pvimap}). 

The lowest degree first integral of the map defined by (\ref{pvimap}), with degree 
pattern $(1,3,3,1)$, is given by 
\beq\label{pvih1} \begin{array}{rcl}
H_1 & =& \big( w_1^2 w_2^2-\de^2(w_1^2+w_2^2)\big)\Big( w_3w_2 +w_0w_1+w_1w_2-w_3w_0+\bar{\nu}\Big)
\\ 
&&
\de^4(w_3w_2+w_0w_1 -w_0w_3) + \bar{c}w_2 w_1 + \bar{a}(w_2+ w_1).
\end{array} 
\eeq  
A nondegenerate Poisson bracket between the coordinates is given by 
$$ 
\{ \, w_n,w_{n+1}\,\} =0, \qquad 
\{ \, w_n,w_{n+2}\,\} = \frac{1}{w_{n+1}^2-\de^2}, 
$$ 
$$ 
\{ \, w_n,w_{n+3}\,\} = - \frac{2(w_n w_{n+1}+w_{n+1}w_{n+2}+w_{n+2}w_{n+3})+\bar{\nu}}{(w_{n+1}^2-\de^2)(w_{n+2}^2-\de^2)}, 
$$
and was derived in \cite{gjtv2} using a discrete Lagrangian structure for (\ref{pvimap}). 
A second independent first integral $H_2$ was given in  \cite{gjtv2}, which is in involution with $H_1$ with respect to this bracket. Here we take the second independent quantity as 
\small 
\beq\label{pvih2} \begin{array}{rl}
H_2 = &
(w_1^2-\de^2)(w_2^2-\de^2)^2\,w_3^2+ (w_1^2-\de^2)^2(w_2^2-\de^2)\,w_0^2
+(2w_1w_2+\bar{\nu})(w_1^2-\de^2)(w_2^2-\de^2)\,w_3w_0 \\ 
& +
\big(2w_1^3w_2^2+\bar{\nu}w_1^2w_2+\bar{c}w_1+\bar{a}
-(2w_1w_2^2+\bar{\nu}w_2)\de^2-w_1\de^4\big)(w_2^2-\de^2)\,w_3  \\
& + 
\big(2w_1^2w_2^3+\bar{\nu}w_1w_2^2+\bar{c}w_2+\bar{a}
-(2w_1^2w_2+\bar{\nu}w_1)\de^2-w_2\de^4\big)(w_1^2-\de^2)\,w_0  \\
&+ 
w_1^4w_2^4+\bar{\nu}w_1^3w_2^3+\bar{c}w_1^2w_2^2+\bar{a}w_1w_2(w_1+w_2) 
\\& 
- 
\Big(\big(w_1^2w_2^2+\bar{\nu}w_1w_2\big)(w_1^2+w_2^2)+\bar{a}(w_1+w_2)  
\Big)\de^2 
\\ & + 
(w_1^2w_2^2+\bar{\nu}w_1w_2-\bar{c})\de^4 -(w_1^2+w_2^2)\de^6;
\end{array} 
\eeq 
\normalsize
so 
the map (\ref{pvimap}) is Liouville integrable. 

The Hamiltonian vector field for the first flow, generated by $H_1$,  
takes the form 
\beq\label{mvolterra2} 
\frac{\rd w_n}{\rd t} = (w_n^2 -\de^2)(w_{n+1}-w_{n-1})
\eeq 
for $n=1,2$, and once again, since the Poisson map (\ref{pvimap}) is compatible with the flow $\{\cdot , H_1\}$, the equation (\ref{mvolterra2}) holds for all 
$n\in\Z$, and  thus the  map and the flow together produce a sequence of functions 
$\big(w_n(t)\big)_{n\in\Z}$ satisfying (\ref{mvolterra2}), which (up to rescaling) is the general form of the modified Volterra lattice equation. If we set 
$\de\to  0$ in 
(\ref{mvolterra2}), then the equation (\ref{mvolterra1}) is recovered, corresponding to the same limit that reproduces 
(\ref{pvmap}) as a degenerate case of (\ref{pvimap}). However, the behaviour of the degenerate map (\ref{pvmap}) is sufficiently different compared with (\ref{pvimap})  e.g.\ with respect to 
singularity structure, that it is worth giving it a separate analysis as we have done here. 

Let us denote a solution of the Volterra lattice (\ref{volterra}) by $\hat{w}_n$. Then the Miura map from the modified Volterra lattice (\ref{mvolterra2}) takes the form 
\beq\label{miuramap} 
\hat{w}_n=(w_{n+1}\mp \de)(w_{n}\pm \de),  
\eeq 
(so there are effectively two maps, with an opposite choice of sign in each factor on the right-hand side above). Moreover, this persists at the level of the maps (\ref{pivmap}) and (\ref{pvimap}), in the following sense. 

\begin{thm} \label{miura} 
Let $w_n$ be a solution of (\ref{pvimap}) with parameters $\bar{a},\bar{c},\bar{\nu}$,  lying on the level set $H_1=\bar{h}_1$, $H_2=\bar{h}_2$ of the first integrals 
(\ref{pvih1}) and (\ref{pvih2}). 
Then for either choice of signs, $\hat{w}_n$ given by the Miura map (\ref{miuramap}) is a solution of (\ref{pivmap}) with parameters 
$$ 
\nu = \bar{\nu}+6\de^2, \qquad b=\bar{c}+4\bar{\nu}\de^2+7\de^4, \qquad a=\bar{h}_1+\bar{c}\de^2+\bar{\nu}\de^4-\de^6.
$$ 
Moreover, on either solution $\hat{w}_n$, 
the values $h_1, h_2$ of the first integrals  (\ref{pivh1}) and  (\ref{pivh2}) for the map  (\ref{pivmap}) 
are given by 
$$ 
h_1=\bar{h}_2+2\de^8, \qquad 
h_2= -\bar{a}^2-\bar{\nu}\bar{h}_2-\bar{c}\bar{h}_1-2\bar{h}_2\de^2
+(\bar{h}_1-\bar{\nu}\bar{c})\de^4-\bar{\nu}\de^8-4\de^{10}. 
$$
\end{thm} 
\begin{prf} The first part of this result is verified by substituting the Miura formula (\ref{miuramap}) directly into (\ref{pivmap}), using (\ref{pvimap}) to eliminate $w_{n+5}$ followed by 
$w_{n+4}$, and then using (\ref{pvih1}) to eliminate $w_{n+3}$ on the level set $H_1=\bar{h}_1$.
After the initial substitution of the Miura map and eliminating, all 
the final results are quadratic in $\de$, so do not depend on the choice of sign in  (\ref{miuramap}).
Similar calculations using the same substitutions in the formulae  (\ref{pivh1}) and  (\ref{pivh2}), together with the expression (\ref{pvih2}) on the level set $H_2=\bar{h}_2$, produce the 
expressions for $h_1,h_2$, which are the corresponding  values of the first integrals for (\ref{pivmap}). 
\end{prf}

We can also make use of a tau function substitution for (P.vi), which has the more complicated structure
\beq\label{pvisigtauplus} 
w_n+\de=\rho_n\, \frac{\sigma_{n+2}\tau_{n}}{\sigma_{n+1}\tau_{n+1}},
\eeq 
\beq\label{pvisigtauminus} 
w_n-\de=\frac{1}{\rho_{n+1}}\, \frac{\sigma_{n}\tau_{n+2}}{\sigma_{n+1}\tau_{n+1}},
\eeq 
with 
$$
\rho_{n+2}=\rho_n. 
$$ 
This implies that 
\beq\label{pvisigma} 
\hat{w}_n^{(+)}=(w_n-\de)(w_{n+1}+\de)=\frac{\sigma_{n}\sigma_{n+3}}{\sigma_{n+1}\sigma_{n+2}}, 
\eeq 
\beq\label{pvitau} 
\hat{w}_n^{(-)}=
(w_n+\de)(w_{n+1}-\de)=\frac{\tau_{n}\tau_{n+3}}{\tau_{n+1}\tau_{n+2}}
\eeq 
are both solutions of (\ref{pivmap}), 
and both sequences $(\sigma_n)$ and $(\tau_n)$ satisfy the same Somos-9 relation, of the form (\ref{s9g2}) (see  Proposition 2.1 in \cite{voltmaps}). Thus the two 
different formulae for the Miura map in  (\ref{miuramap}) can be regarded as defining a B\"acklund transformation for the discrete equation (\ref{pivmap}) with parameter $\de$, since given $\hat{w}_n^{(-)}$ and a solution $w_n$ of (\ref{pvimap}), a new solution $\hat{w}_n^{(+)}$ of 
(\ref{pivmap}) is generated by taking 
$$ 
\hat{w}_n^{(+)} = \hat{w}_n^{(-)}+2\de (w_{n+1}-w_n).
$$

\section{Conclusion}  

We have shown that the integrable maps (P.iv), (P.v)  and (P.vi) from \cite{gjtv2} are closely related to one another, via Miura-type transformations, and 
they provide genus two solutions of Volterra and modified Volterra lattices, respectively. So far we do not have a complete understanding of what the relations between 
these maps mean  
geometrically, particularly from the Poisson and algebro-geometric points of view. 
However, since the construction of the integrable maps ${\cal V}_g$ presented in \cite{voltmaps} is 
valid for any $g\geq 1$, this strongly suggests that (P.v) and (P.vi) should each be the $g=2$ members of a family of maps defined for any $g$. In the elliptic case ($g=1$) 
we have constructed elliptic solutions of the modified Volterra and Volterra lattices, and showed how they are linked by the Miura transformation, essentially recovering the 
solutions found in \cite{yan}, which can be interpreted in terms of integrable maps in the plane (QRT type). The complete description of these results, together with 
the proposed extension to families of maps for all $g\geq 1$, is planned for future work.

\subsection*{Acknowledgements}

The research of ANWH was supported by Fellowship EP/M004333/1 from the
Engineering \& Physical Sciences Research Council, UK, extended by EP/V520718/1 
COVID 19 Grant Extension Allocation University of Kent, and the grant IEC\textbackslash R3\textbackslash 193024 from the
Royal Society; he is also grateful to the School of Mathematics and Statistics, University of New South Wales, for
hosting him during 2017-2019 as a Visiting Professorial Fellow with funding from the Distinguished Researcher
Visitor Scheme, and to Wolfgang Schief, who provided additional support during his time in Sydney. ANWH also thanks DICATAM 
for supporting his visit to Brescia in November 2022. FZ acknowledges the support of Universit\`a di Brescia,  GNFM-INdAM and INFN, Sezione di Milano-Bicocca, Gr. IV - Mathematical Methods in NonLinear Physics (Milano, Italy).

\label{lastpage}
\end{document}